\documentclass[11pt,aps,showpacs,nofootinbib,prd,aps,epsf,floats,axodraw,amsmath,amssymb,amsfonts]{revtex4}
\usepackage{amsmath, amssymb}

\newcommand{\mathsym}[1]{{}}

\usepackage{graphicx}
\usepackage{amsmath}
\usepackage{amssymb}
\usepackage{bm}
\newcommand{\be}{\begin{equation}}
\newcommand{\ee}{\end{equation}}
\newcommand{\bea}{\begin{eqnarray}}
\newcommand{\eea}{\end{eqnarray}}

\newcommand{\rem}[1]{}
\newsavebox{\PSLASH}
 \sbox{\PSLASH}{$p$\hspace{-1.8mm}/}
 
\renewcommand{\theequation}{\thesection.\arabic{equation}}
\newcounter{saveeqn}
\newcommand{\add}{\addtocounter{equation}{1}}
\newcommand{\alpheqn}{\setcounter{saveeqn}{\value{equation}}%
\setcounter{equation}{0}%
\renewcommand{\theequation}{\mbox{\thesection.\arabic{saveeqn}{\alph{equation}}}}}
\newcommand{\reseteqn}{\setcounter{equation}{\value{saveeqn}}%
\renewcommand{\theequation}{\thesection.\arabic{equation}}}

 \newsavebox{\notrightarrow}
 \sbox{\notrightarrow}{$\to$\hspace{-4mm}/}
 
 \newsavebox{\PARTIALSLASH}
 \sbox{\PARTIALSLASH}{$\partial$\hspace{-1.6mm}/}
 
 \newsavebox{\ASLASH}
 \sbox{\ASLASH}{$A$\hspace{-2.1mm}/}
 
 \newsavebox{\KSLASH}
 \sbox{\KSLASH}{$k$\hspace{-1.8mm}/}
 
 \newsavebox{\LSLASH}
 \sbox{\LSLASH}{$\ell$\hspace{-1.8mm}/}
 
 \newsavebox{\QSLASH}
 \sbox{\QSLASH}{$q$\hspace{-1.8mm}/}
 
 \newsavebox{\DSLASH}
 \sbox{\DSLASH}{$D$\hspace{-2.2mm}/}
 
 \newsavebox{\DbfSLASH}
 \sbox{\DbfSLASH}{${\mathbf D}$\hspace{-2.8mm}/}
 
 \newsavebox{\DELVECRIGHT}
 \sbox{\DELVECRIGHT}{$\stackrel{\rightarrow}{\partial}$}
 
 \newcommand{\blue}{\IfColor{\textCadetBlue}{}}
\newcommand{\black}{\IfColor{\textBlack}{}}
\newcommand{\red}{\IfColor{\textRed}{}}
\newcommand{\green}{\IfColor{\textOliveGreen}{}}
\newcommand{\lila}{\IfColor{\textRedViolet}{}}

\begin{document}
\title{Study of the photon's pole structure in the noncommutative Schwinger model}
\author{M. Ghasemkhani$^{1,2}$}\email{ghasemkhani@ipm.ir}
\affiliation{$^1$Department of Physics, Shahid Beheshti University, G. C., Evin, Tehran 19839, Iran\\
$^{2}$School of Physics, Institute for Research in Fundamental Sciences (IPM),\\
 P.O.Box 19395-5531, Tehran, Iran}
\begin{abstract}
\noindent The photon self-energy of the noncommutative Schwinger
model at two- and three-loop order is analyzed. It is shown that the
mass spectrum of the model does not receive any correction from
noncommutativity parameter ($\theta$) at these orders. Also it
remains unchanged to all orders. The exact one-loop effective action
for the photon is also calculated.
\end{abstract}
 \maketitle
\section{Introduction}\label{introduction}
\noindent The idea of noncommutative quantum field theory
originates from the 1940s, when it was applied to cure the divergencies
in quantum field theory before the renormalization approach was born
\cite{snyder}. It was demonstrated that the
divergencies were not removed \cite{yang}. Later on, it was shown in
\cite{witten} that the noncommutative quantum field theory describes
effectively the low energy limit of the string theory on a
noncommutative manifold. In the simplest case, the description of
the noncommutative space-time is given by a constant parameter,
$\theta^{\mu\nu}$, of which the space-space (-time) components
correspond to the magnetic (electric) field. The space-time
noncommutative field theories suffer from the unitarity violation of
the S-matrix \cite{mehen} while the space-space noncommutative field
theories face another obstacle, mixing of ultraviolet and infrared
singularities \cite{minwalla}. The problem of the non-unitary
S-matrix was studied in \cite{fredenhagen-1,fredenhagen-2,fredenhagen-3} but these works include some
inconsistencies.\\ In fact, space-time noncommutativity leads to the
higher orders of time derivatives of the fields in the Lagrangian
which make the quantization procedure of the theory different from
that of the commutative counterpart. For example in
\cite{ghasemkhani}, the perturbative quantization of the
noncommutative QED in 1+1 dimensions has been analyzed up to
${\cal{O}}(\theta^{3})$.
\par In the present work, the
noncommutative two-dimensional QED with massless fermions in
Euclidean space $(x_{2}\equiv it)$ is considered. The purpose of
this paper is to concentrate on the mass spectrum of the theory at
higher loops. The commutative counterpart of this model, Schwinger
model, was studied in \cite{schwinger} where it was shown that the
photon in two dimensions acquires dynamical mass, arising from the
loop effect, without gauge symmetry breaking. The mass spectrum of
the Schwinger model contains a free boson with a mass proportional
to the dimensionful coupling constant. Fermions disappear from the
physical states due to the linearity of the potential that is
similar to the quark confinement potential in quantum
chromodynamics (QCD). Hence, Schwinger model can be a toy model to
understand the quark confinement. The extension of the Schwinger
model to the noncommutative version as regards different aspects has been
addressed in
\cite{rahaman1,rahaman2,harikumar,ghasemkhani,ardalan,armoni,petrov}.
 Here, we focus on the dynamical
mass generation in the noncommutative space.
\par
This paper is organized as follows: in Sect. II, we introduce the
noncommutative Schwinger model in the light-cone coordinates in
order to simplify our calculations. In Sect. III, to obtain the mass
spectrum of the theory at two- and three-loop order, the photon self-energy is studied. Using the explicit representation of the Dirac
$\gamma$-matrices provides a straightforward method to compute the
trace of the complicated fermionic loops. Then it is shown that the
noncommutativity does not affect the Schwinger mass at these levels.
The computations of Sect. III are extended to all orders
 in Sect. IV where the exact mass spectrum is also obtained.
In Sect. V, we demonstrate that the noncommutative one-loop effective
action for the photon is exactly the same as the commutative
counterpart. Finally, Sect. VI is devoted to the concluding remarks.
\par\noindent
\section{Noncommutative Schwinger model in the light-cone coordinates}
\setcounter{equation}{0} \par\noindent The Lagrangian of the
noncommutative Schwinger model can be obtained from its commutative
counterpart by replacing the ordinary product with the star-product,
which is defined as follows
\begin{eqnarray}
f(x)\star g(x)\equiv
\exp\left(\frac{i\theta_{\mu\nu}}{2}\frac{\partial}{\partial
a_{\mu}}\frac{\partial}{\partial b_{\nu}}\right)\
f(x+a)g(x+b)\Bigg|_{a=b=0},
\end{eqnarray}
where $\theta_{\mu\nu}$ is an antisymmetric constant matrix related
to the noncommutative structure of the space-time. In
two-dimensional space-time, $\theta_{\mu\nu}$ can be written as the
antisymmetric tensor $\epsilon_{\mu\nu}$ which preserves the Lorentz
symmetry, namely
\begin{eqnarray}
[x_{\mu},x_{\nu}]=\theta\epsilon_{\mu\nu}.
\end{eqnarray}
To avoid the unitarity problem in the noncommutative space-time
field theories, we use the Euclidean signature throughout this
paper. The Lagrangian of the two-dimensional noncommutative massless
QED is given by
\begin{eqnarray}\label{G2}
{\cal{L}}&=&-i\bar{\psi}\star\gamma_{\mu}\partial^{\mu} \psi+e
\bar{\psi}\star \gamma_{\mu}A^{\mu}\star
\psi+\frac{1}{4}F_{\mu\nu}\star
F^{\mu\nu}+\frac{1}{2}(\partial_{\mu} A^{\mu})\star (\partial_{\nu}
A^{\nu})\nonumber\\&-&\partial_{\mu}\bar c\star
(\partial^{\mu}c-ie[A^{\mu},c]_{\star}),
\end{eqnarray}
where $F_{\mu\nu}$ is defined as
\begin{eqnarray}
F_{\mu\nu}= \partial_{\mu} A_{\nu}-\partial_{\nu} A_{\mu}+ie
[A_{\mu},A_{\nu}]_{\star},
\end{eqnarray}
with $[A_{\mu},A_{\nu}]_{\star} = A_{\mu}\star A_{\nu}-A_{\nu}\star
A_{\mu}$. One of the useful properties of the two-dimensional space
is that our calculations in the light-cone coordinates,
$x_{\pm}=x_{1}\pm ix_{2}$, are simplified significantly. The
Lagrangian (\ref{G2}) in the light-cone gauge, $A_{-}=0$, has the
following form:
\begin{eqnarray}
{\cal L }=
\frac{i}{2}\bar\psi\star(\gamma_{+}\partial_{-}+\gamma_{-}\partial_{+})\psi+\frac{e}{2}
\bar\psi\star\gamma_{-}
A_{+}\star\psi-\frac{1}{2}(\partial_{-}A_{+})\star
(\partial_{-}A_{+}),
\end{eqnarray}
where $\gamma_{\pm}=\gamma_{1}\pm i\gamma_{2}$ and $A_{\pm}=A_{1}\pm
iA_{2}$.\\ In this particular gauge, the non-linear term in the
field strength tensor is removed. Therefore, the photon self-interaction parts, three- and four-photon interaction vertices, are
eliminated and the ghost fields are decoupled from the theory. The
resulting Feynman rules are shown in Fig. 1.
\begin{figure}[h]
\vspace{0.5cm} \centering
  \includegraphics[width=8cm,height=6cm]{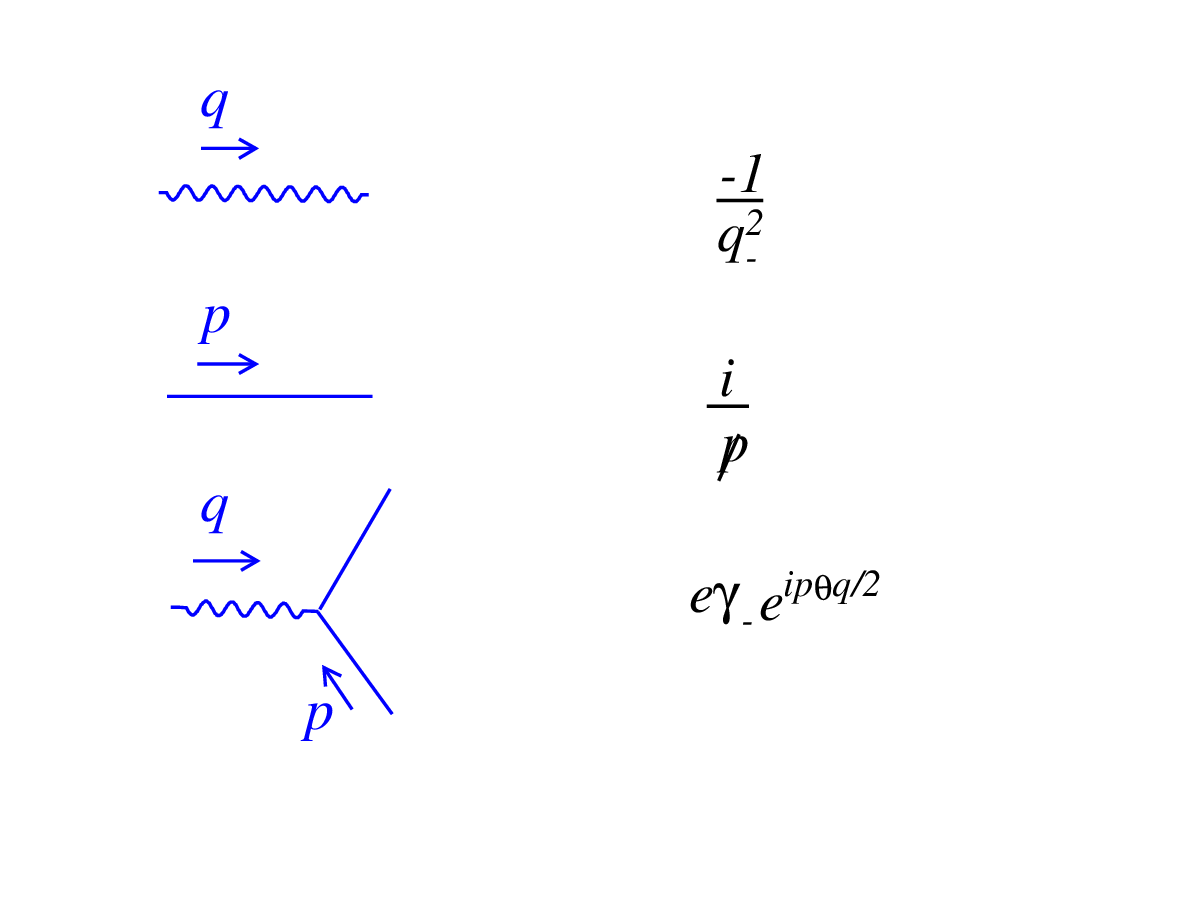}
  \caption{Feynman rules for noncommutative Schwinger model in the light-cone gauge
}\label{FinalFeynmanRules}
\end{figure}
\\
 Note that only $\gamma_{-}$ appears in the fermion-photon vertex.
\section{Two- and three-loop noncommutative correction to the Schwinger mass  }
\setcounter{equation}{0}
\par\noindent
As was mentioned before, Schwinger showed that the photon in two
dimensions acquires dynamical mass, $\mu=\frac{e}{\sqrt{\pi}}$. This
mass generation originates from the presence of a special
singularity in the scalar vacuum polarization at one-loop order.
Using the non-perturbative method shows that the obtained mass does
not receive any correction from loops at higher orders
\cite{schwinger,abdalla}. The noncommutative extension of this kind
of mass generation at one-loop level was discussed in \cite{ardalan}
where it was proved that the Schwinger mass gets no noncommutative
correction in this order. Higher-loop contributions, e.g. two- and
three-loop contributions,
 have been pointed out in \cite{armoni} without explicit
computation of the loop
integrals.\\
At two-loop order, there is only one diagram with $\theta$-dependent
phase factor, but three-loop order includes three $\theta$-dependent
graphs. It is shown that the two- and three-loop computations are
very similar. However, the analysis of the relevant three-loop
graphs is a bit more complicated than that of the
two-loop graph. \\
 The general structure of the exact photon
propagator\footnote{Here, we work in Feynman gauge.} in two-dimensional noncommutative space is the same as its commutative
counterpart \cite{ardalan}, namely
\begin{eqnarray}
D^{\mu\nu}(q)=-\frac{\delta^{\mu\nu}}{q^{2}[1+\Pi(q^{2})]},
\end{eqnarray}
where the scalar vacuum polarization, $\Pi(q^{2})$, is related to
its tensor form via the following:
\begin{eqnarray}
\Pi^{\mu\nu}=(q^{2}\delta^{\mu\nu}-q^{\mu}q^{\nu})\Pi(q^{2}),
\end{eqnarray}
where $\Pi(q^{2})$ includes the commutative and noncommutative
parts. The pole structure is obtained from the following limit:
\begin{eqnarray}\label{spectrum}
\lim\limits_{q^{2}\rightarrow
0}q^{2}\Pi(q^{2},e^{2},\theta)=\lim\limits_{q^{2}\rightarrow
0}q^{2}\Pi_{\mbox\tiny{c}}(q^{2},e^{2})+\lim\limits_{q^{2}\rightarrow
0}q^{2}\Pi_{\mbox\tiny{nc}}(q^{2},e^{2},\theta),
\end{eqnarray}
with fixed $\theta$. The first term yields the exact commutative
Schwinger mass with $\Pi(q^{2},e^{2})=\frac{e^{2}}{\pi q^{2}}$
 and the second term gives the
noncommutative corrections to it. In the present section, we
concentrate on the analysis of the second term in (\ref{spectrum})
at two- and three-loop level.
\subsection{Two-loop noncommutative correction}
\noindent
 Two-loop order contains only one $\theta$-dependent
diagram which is shown in Fig. 2. Here and in all figures of the paper, it is notable that a small circle oriented with pink arrows indicates a twist and does not show a fermionic loop. The Feynman form related to Fig. 2 is given
by
\begin{eqnarray}
\Pi_{\mu\nu}^{(2)}|_{nc}=e^{4}\int\frac{d^{2}p}{(2\pi)^{2}}\frac{d^{2}k}{(2\pi)^{2}}
\frac{1}{k^{2}} e^{-ik\theta q}~tr\bigg(\gamma_{\mu}
\frac{1}{(\displaystyle{\not}q+\displaystyle{\not}p)}
\gamma^{\rho}\frac{1}{(\displaystyle{\not}q+\displaystyle{\not}p+\displaystyle{\not}k)}
\gamma_{\nu}\frac{1}{(\displaystyle{\not}p+\displaystyle{\not}k)}\gamma_{\rho}
\frac{1}{\displaystyle{\not}p}\bigg),
\end{eqnarray}
which in the light-cone coordinates leads to the following
\begin{eqnarray}\label{2loop self}
\Pi_{--}^{(2)}|_{nc}=
e^{4}\int\frac{dp_{-}dp_{+}}{(2\pi)^{2}}\frac{dk_{+}dk_{-}}{(2\pi)^{2}}
\frac{g^{+-}e^{-ik\theta q}{\cal
N}}{k^{2}_{-}(q+p)^{2}(q+p+k)^{2}(p+k)^{2}p^{2}},
\end{eqnarray}
where
\begin{eqnarray}\label{2loopnum}
{\cal N}=tr(\gamma_{-}
(\displaystyle{\not}q+\displaystyle{\not}p)\gamma_{-}(\displaystyle{\not}q+
\displaystyle{\not}p+\displaystyle{\not}k)
\gamma_{-}(\displaystyle{\not}p+\displaystyle{\not}k)\gamma_{-}
\displaystyle{\not}p ),
\end{eqnarray}
 and
$k\theta
q=\theta_{\mu\nu}k_{\mu}q_{\nu}=\frac{i\theta}{2}(k_{+}q_{-}-k_{-}q_{+})$.
Using the explicit matrix form of $\gamma_{-}$ is useful to find the
trace of the fermionic loop in a simple way (see Appendix A for more
details). Therefore, the value of ${\cal N}$ is obtained as
\begin{eqnarray}\label{value of 2loop }
{\cal N}= 2^{4}~(p+q)_{-}(p+q+k)_{-}(p+k)_{-}p_{-}.
\end{eqnarray}
Putting (\ref{value of 2loop }) in (\ref{2loop self}), we have
\begin{eqnarray}
\Pi_{--}^{(2)}|_{nc}=8
e^{4}\int\frac{dp_{-}dp_{+}}{(2\pi)^{2}}\frac{dk_{+}dk_{-}}{(2\pi)^{2}}
\frac{e^{-ik\theta
q}(p+q)_{-}(p+q+k)_{-}(p+k)_{-}p_{-}}{k^{2}_{-}(q+p)^{2}(q+p+k)^{2}(p+k)^{2}p^{2}},
\end{eqnarray}
that is rewritten as
\begin{eqnarray}\label{finaltwoloop}
\Pi_{--}^{(2)}|_{nc}=
8e^{4}\int\frac{dk_{+}}{2\pi}\frac{dk_{-}}{2\pi}
\frac{1}{k_{-}^{2}}~e^{-ik\theta q}{\cal E},
\end{eqnarray}
with
\begin{eqnarray}\label{2loop p-integral}
{\cal E}=\int
\frac{dp_{-}}{2\pi}\frac{dp_{+}}{2\pi}\frac{1}{(p+q)_{+}(p+q+k)_{+}
(p+k)_{+}p_{+}}.
\end{eqnarray}
 The produced phase factor in (\ref{finaltwoloop}) is independent of the fermionic-loop
 momentum; hence the integral over $p$ can be evaluated separately.
 \begin{figure}[h]
  \vspace{-0.4cm}
  \includegraphics[width=10cm,height=6cm]{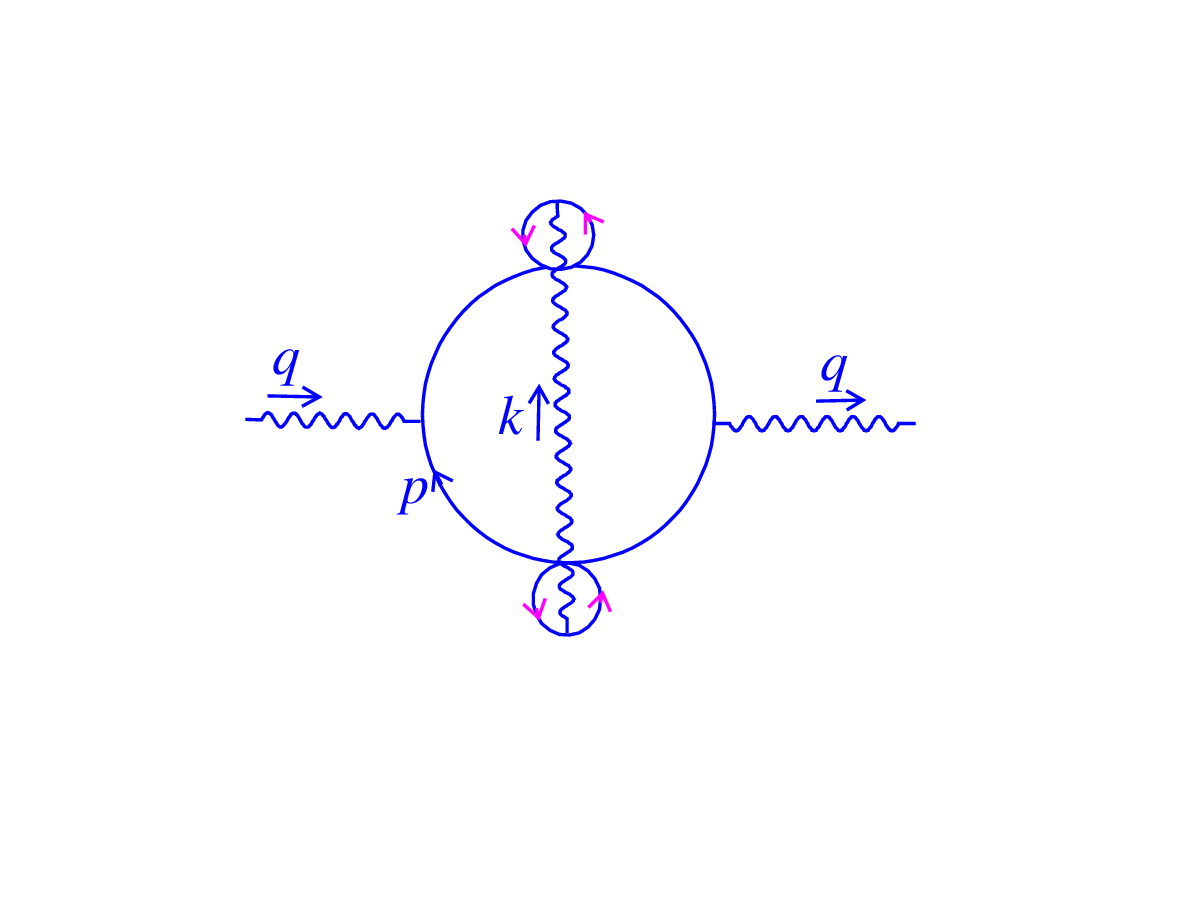}
  \vspace{-1cm}\centering
  \caption{Relevant two-loop diagram
}\label{2LOOP}
\end{figure}
 \\To simplify (\ref{2loop p-integral}), we decompose the fraction into partial
fractions to reduce the degree of the denominator. The first step of
the decomposition results in
\begin{eqnarray}
{\cal E}=\int \frac{dp_{-}}{2\pi}\frac{dp_{+}}{2\pi}~
\frac{1}{k^{2}_{+}}\bigg[\frac{1}{(p+q)_{+}}-\frac{1}{(p+q+k)_{+}}\bigg]\bigg[\frac{1}{p_{+}}-\frac{1}{(p+k)_{+}}\bigg].
\end{eqnarray}
Performing the complete decomposition produces the final expression
as
\begin{eqnarray}\label{simplified result of two loop}
{\cal E}&=&\int\frac{dp_{-}}{2\pi}\frac{
dp_{+}}{2\pi}~\frac{1}{k^{2}_{+}} ~\bigg\{
\frac{1}{q_{+}}\bigg[\frac{1}{p_{+}}-\frac{1}{(p+q)_{+}}\bigg]
-\frac{1}{(k-q)_{+}}\bigg[\frac{1}{(p+q)_{+}}-\frac{1}{(p+k)_{+}}\bigg]
\nonumber\\&-&\frac{1}{(k+q)_{+}}\bigg[\frac{1}{p_{+}}-\frac{1}{(p+q+k)_{+}}\bigg]
+\frac{1}{q_{+}}\bigg[\frac{1}{(p+k)_{+}}-\frac{1}{(p+q+k)_{+}}\bigg]\bigg\}.
\end{eqnarray}
According to the complex form of Green's theorem mentioned in
\cite{schaum}, it is deduced that the $p$-integrals in each pairs
separated in the parentheses vanish, namely ${\cal E}=0$.
 Hence
 \begin{eqnarray}\label{2loopresult}
\Pi_{--}^{(2)}|_{nc}=0.
 \end{eqnarray}
If we use the electron mass as an infrared regulator, the obtained
result remains unchanged.
 The detailed calculations with infrared regulator will
be presented in Appendix B.\\
 According to (\ref{spectrum}), the
commutative Schwinger mass remains free from the noncommutative
correction at two-loop order. In what follows, this calculation will
be extended to three-loop level of the quantum corrections.
\subsection{Three-loop noncommutative correction}
\noindent
At three-loop order, unlike the two-loop case, there is more than one graph with
$\theta$-dependent phase factor. Some of these graphs have been represented in Fig. 3.
 The contributions related to the
graphs (a), (b) and (c) of Fig. 3 can be expressed as the following,
respectively
\begin{eqnarray}\label{covariantcoordinates}
\Pi_{\mu\nu}^{(3)}|_{nc}&=&e^{6}\int\frac{d^{2}p}{(2\pi)^{2}}\frac{d^{2}k}{(2\pi)^{2}}\frac{d^{2}\ell}{(2\pi)^{2}}
\frac{1}{k^{2}}\frac{1}{\ell^{2}}\nonumber\\&\times&\bigg\{
e^{-i(k\theta \ell+k\theta q+\ell\theta q)}tr\bigg(\gamma_{\mu}
\frac{1}{(\displaystyle{\not}q+\displaystyle{\not}p)}\gamma^{\rho}\frac{1}{(\displaystyle{\not}q+\displaystyle{\not}p+\displaystyle{\not}\ell)}\gamma^{\lambda}\frac{1}{(\displaystyle{\not}q+\displaystyle{\not}p+\displaystyle{\not}\ell+\displaystyle{\not}k)}
\gamma_{\nu}\frac{1}{(\displaystyle{\not}p+\displaystyle{\not}\ell+\displaystyle{\not}k
)
}\gamma_{\rho}\frac{1}{(\displaystyle{\not}p+\displaystyle{\not}k)}\gamma_{\lambda}
\frac{1}{\displaystyle{\not}p}\bigg)\nonumber\\&+& e^{-i(k\theta
q+\ell\theta q)}tr\bigg(\gamma_{\mu}
\frac{1}{(\displaystyle{\not}q+\displaystyle{\not}p)}\gamma^{\rho}\frac{1}{(\displaystyle{\not}q+\displaystyle{\not}p+
\displaystyle{\not}\ell)}\gamma^{\lambda}\frac{1}{(\displaystyle{\not}q+\displaystyle{\not}p+\displaystyle{\not}\ell+\displaystyle{\not}k)}
\gamma_{\nu}\frac{1}{(\displaystyle{\not}p+\displaystyle{\not}\ell+\displaystyle{\not}k)}\gamma_{\lambda}\frac{1}{(\displaystyle{\not}p+
\displaystyle{\not}\ell)}\gamma_{\rho}
\frac{1}{\displaystyle{\not}p}\bigg)\bigg\}\nonumber\\&+&
e^{6}\int\frac{d^{2}p}{(2\pi)^{2}}\frac{d^{2}k}{(2\pi)^{2}} \frac{
e^{-ik\theta q}}{k^{4}}~tr\bigg(\gamma_{\mu}
\frac{1}{(\displaystyle{\not}q+\displaystyle{\not}p)}
\gamma_{\rho}\frac{1}{(\displaystyle{\not}q+\displaystyle{\not}p+\displaystyle{\not}k)}
\gamma_{\nu}\frac{1}{(\displaystyle{\not}p+\displaystyle{\not}k)}\gamma_{\sigma}
\frac{1}{\displaystyle{\not}p}\bigg)
\nonumber\\&\times&\int\frac{d^{2}\ell}{(2\pi)^{2}}~tr\bigg(\gamma^{\rho}\frac{1}
{\displaystyle{\not}\ell}\gamma^{\sigma}\frac{1}{(\displaystyle{\not}\ell+\displaystyle{\not}k)}\bigg)+\cdots.
\end{eqnarray}
Here dots refer to the other diagrams that appear in this order.
Rewriting (\ref{covariantcoordinates}) in the light-cone
coordinates, we obtain
\begin{eqnarray}\label{threeloop}
\Pi_{--}^{(3)}|_{nc}&=&
e^{6}\int\frac{dp_{-}dp_{+}}{(2\pi)^{2}}\frac{dk_{+}dk_{-}}{(2\pi)^{2}}\frac{d\ell_{+}d\ell_{-}}{(2\pi)^{2}}
\frac{g^{+-}g^{+-}}{k^{2}_{-}\ell^{2}_{-}(q+p)^{2}(q+p+\ell)^{2}(q+p+\ell+k)^{2}(p+\ell+k)^{2}p^{2}}\nonumber\\
&\times &\bigg[\frac{{\cal N}_{a}e^{-i(k\theta\ell+k\theta
q+\ell\theta q)}}{(p+k)^{2}}+
\frac{{\cal N}_{b}e^{-i(k\theta q+\ell\theta q)}}{(p+\ell)^{2}}\bigg]\nonumber\\
&+&
e^{6}\int\frac{dp_{-}dp_{+}}{(2\pi)^{2}}\frac{dk_{+}dk_{-}}{(2\pi)^{2}}
\frac{{\cal N}_{c}~e^{-ik\theta
q}g^{+-}g^{+-}}{k^{4}_{-}(q+p)^{2}(q+p+k)^{2}(p+k)^{2}p^{2}}
\int\frac{d\ell_{+}d\ell_{-}}{(2\pi)^{2}}\frac{tr(\gamma_{-}\displaystyle{\not}\ell\gamma_{-}
(\displaystyle{\not}\ell+\displaystyle{\not}k))}{\ell^{2}(\ell+k)^{2}}+\cdots,
\nonumber\\
\end{eqnarray}
where
\begin{eqnarray}\label{numerator}
&&{\cal N}_{a}=tr\bigg(\gamma_{-}
(\displaystyle{\not}q+\displaystyle{\not}p)\gamma_{-}(\displaystyle{\not}q+
\displaystyle{\not}p+\displaystyle{\not}\ell)\gamma_{-}(\displaystyle{\not}q+\displaystyle{\not}p+\displaystyle{\not}\ell+\displaystyle{\not}k)
\gamma_{-}(\displaystyle{\not}p+\displaystyle{\not}\ell+\displaystyle{\not}k)\gamma_{-}(\displaystyle{\not}p+\displaystyle{\not}k)\gamma_{-}
\displaystyle{\not}p \bigg)
\nonumber\\
 &&{\cal N}_{b}=tr\bigg(\gamma_{-}
(\displaystyle{\not}q+\displaystyle{\not}p)\gamma_{-}(\displaystyle{\not}q+
\displaystyle{\not}p+\displaystyle{\not}\ell)\gamma_{-}(\displaystyle{\not}q+\displaystyle{\not}p+\displaystyle{\not}\ell+\displaystyle{\not}k)
\gamma_{-}(\displaystyle{\not}p+\displaystyle{\not}\ell+\displaystyle{\not}k)\gamma_{-}(\displaystyle{\not}p+\displaystyle{\not}\ell)\gamma_{-}
\displaystyle{\not}p \bigg),\nonumber\\ &&
 {\cal N}_{c}=tr\bigg(\gamma_{-}(\displaystyle{\not}q+\displaystyle{\not}p)\gamma_{-}(\displaystyle{\not}q+\displaystyle{\not}p+\displaystyle{\not}k)
 \gamma_{-}(\displaystyle{\not}p+\displaystyle{\not}k)\gamma_{-}\displaystyle{\not}p\bigg).
\end{eqnarray}
\begin{figure}[h]
\vspace{0.04cm} \centering
  \includegraphics[width=12cm,height=6cm]{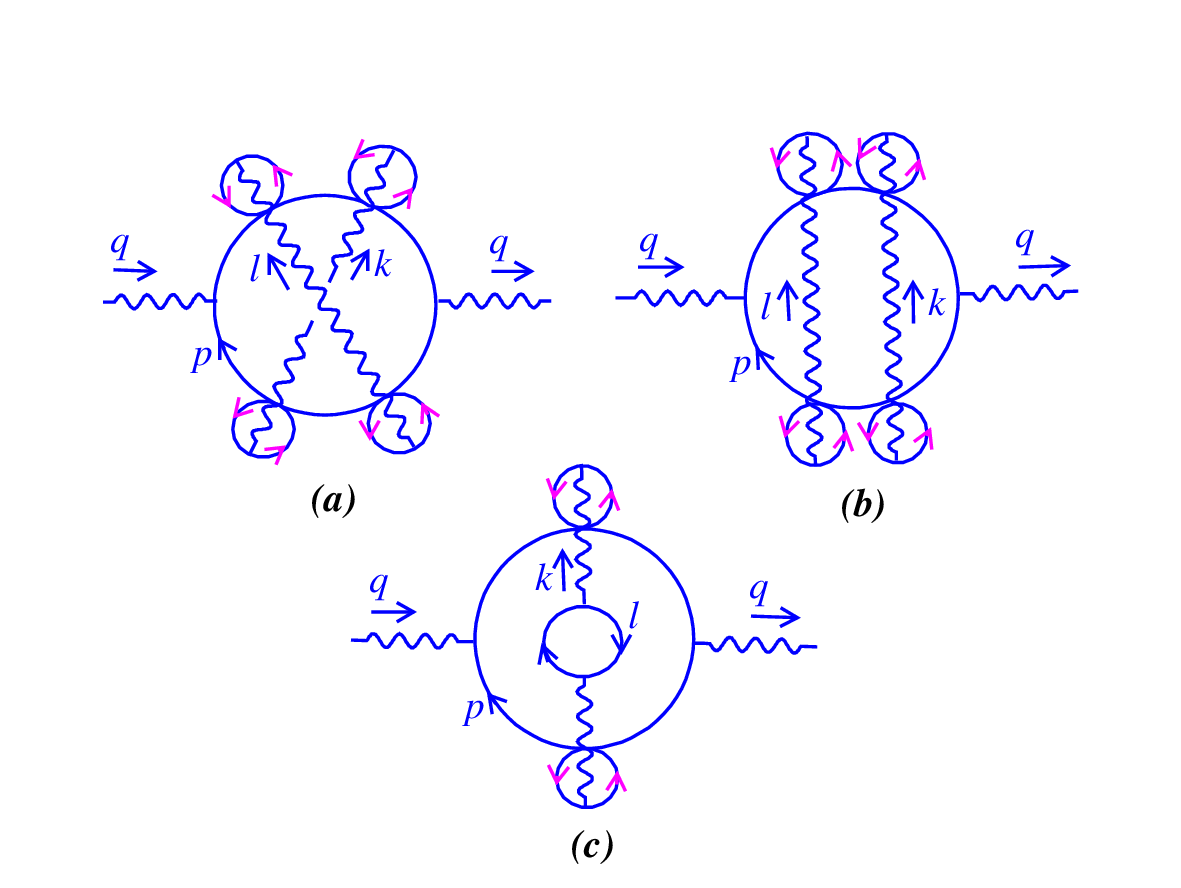}
  \caption{Some of the relevant three-loop diagrams
}\label{nonplanar}
\end{figure}
\\
Having applied the relations mentioned in Appendix A, the
explicit forms of the
 quantities ${\cal N}_{a}$ , ${\cal
N}_{b}$ , and ${\cal N}_{c}$ are given by
\begin{eqnarray}
&&{\cal N}_{a}=
2^{6}~(p+q)_{-}(p+q+\ell)_{-}(p+q+\ell+k)_{-}(p+\ell+k)_{-}(p+k)_{-}p_{-},\nonumber\\&&
{\cal N}_{b}=
2^{6}~(p+q)_{-}(p+q+\ell)_{-}(p+q+\ell+k)_{-}(p+\ell+k)_{-}(p+\ell)_{-}p_{-},\nonumber\\&&
{\cal N}_{c}=2^{4}(p+q)_{-}(p+q+k)_{-}(p+k)_{-}p_{-}.
\end{eqnarray}
Plugging them in (\ref{threeloop}), we have
\begin{eqnarray}\label{finalthreeloop}
\Pi_{--}^{(3)}|_{nc}&=&
16e^{6}\int\frac{dp_{-}dp_{+}}{(2\pi)^{2}}\frac{dk_{+}dk_{-}}{(2\pi)^{2}}\frac{d\ell_{+}d\ell_{-}}{(2\pi)^{2}}
\frac{(p+q)_{-}(p+q+\ell)_{-}(p+q+\ell+k)_{-}(p+\ell+k)_{-}p_{-}}{k^{2}_{-}\ell^{2}_{-}(q+p)^{2}(q+p+\ell)^{2}(q+p+\ell+k)^{2}(p+\ell+k)^{2}p^{2}}\nonumber\\
&\times &\bigg[\frac{e^{-i(k\theta\ell+k\theta q+\ell\theta
q)}(p+k)_{-}}{(p+k)^{2}}+ \frac{e^{-i(k\theta q+\ell\theta
q)}(p+\ell)_{-}}{(p+\ell)^{2}}\bigg] \nonumber\\&+& 4e^{6}
\int\frac{dp_{-}dp_{+}}{(2\pi)^{2}}\frac{dk_{+}dk_{-}}{(2\pi)^{2}}
\frac{e^{-ik\theta
q}(p+q)_{-}(p+q+k)_{-}(p+k)_{-}p_{-}}{k^{4}_{-}(q+p)^{2}(q+p+k)^{2}(p+k)^{2}p^{2}}
\nonumber\\&\times&\int\frac{d\ell_{+}d\ell_{-}}{(2\pi)^{2}}\frac{tr(\gamma_{-}\displaystyle{\not}\ell\gamma_{-}
(\displaystyle{\not}\ell+\displaystyle{\not}k))}{\ell^{2}(\ell+k)^{2}}+\cdots.
\end{eqnarray}
As we see the phase factors appearing in (\ref{finalthreeloop}) ,
similar to the two-loop calculation, are independent of the
fermionic-loop momentum. It can be shown that the other graphs, which appeared at three-loop level, also have fermionic-loop momentum-independent noncommutative phase factor. In fact, this property remains true for all of the diagrams at any order \cite{ghasemkhani-2}. Consequently, the $p$-integrals are calculated independently. Consider the first term of
 (\ref{finalthreeloop})
\begin{eqnarray}
&&\Pi_{--}^{(3,a)}|_{nc}=
16e^{6}\int\frac{dk_{+}dk_{-}}{(2\pi)^{2}}\frac{d\ell_{+}d\ell_{-}}{(2\pi)^{2}}
~\frac{1}{k_{-}^{2}\ell_{-}^{2}}~e^{-i(k\theta\ell+k\theta
q+\ell\theta q)}{\cal F}+\cdots,
\end{eqnarray}
where
\begin{eqnarray}\label{f-term}
{\cal F}=\int\frac{dp_{-}}{2\pi}
\frac{dp_{+}}{2\pi}\frac{1}{(p+q)_{+}(p+q+\ell)_{+}(p+q+\ell+k)_{+}
(p+\ell+k)_{+}(p+k)_{+}p_{+}}.
\end{eqnarray}
We use the decomposition method to simplify (\ref{f-term}).  Using
the
 decomposition method at the first step leads to
\begin{eqnarray}
{\cal F}&=&\int
\frac{dp_{-}}{2\pi}\frac{dp_{+}}{2\pi}\frac{1}{(kq\ell)_{+}
}\bigg[\frac{1}{(p+q)_{+}}-\frac{1}{(p+q+\ell)_{+}}\bigg]
\bigg[\frac{1}{(p+k+\ell)_{+}}-\frac{1}{(p+q+\ell+k)_{+}}\bigg]
\nonumber\\&\times&\bigg[\frac{1}{p_{+}}-\frac{1}{(p+k)_{+}}\bigg],
\end{eqnarray}
and in the second step, we find
\begin{eqnarray}\label{second step}
{\cal F}&=&\int
\frac{dp_{-}}{2\pi}\frac{dp_{+}}{2\pi}\frac{1}{(kq\ell)_{+} }\bigg\{
\frac{1}{(\ell+k-q)_{+}}
\bigg[\frac{1}{(p+q)_{+}}-\frac{1}{(p+q+k)_{+}}\bigg]
\nonumber\\&
-&\frac{1}{(\ell+k)_{+}}\bigg[\frac{1}{(p+q)_{+}}-\frac{1}{(p+q+\ell+k)_{+}}\bigg]
-\frac{1}{(k-q)_{+}}\bigg[\frac{1}{(p+q+\ell)_{+}}-\frac{1}{(p+\ell+k)_{+}}\bigg]
\nonumber\\&+&\frac{1}{k_{+}}\bigg[\frac{1}{(p+q+\ell)_{+}}-\frac{1}{(p+q+\ell+k)_{+}}\bigg]
\bigg\}\bigg[\frac{1}{p_{+}}-\frac{1}{(p+k)_{+}}\bigg].
\end{eqnarray}
After some algebraic manipulations, (\ref{second step}) is reduced
to the following expression
\begin{eqnarray}\label{final f-term}
{\cal F}&=&\int\frac{dp_{-}}{2\pi}
\frac{dp_{+}}{2\pi}\bigg\{\frac{1}{(k
q)_{+}}\frac{1}{(k+\ell)_{+}}\frac{1}{(q+\ell)_{+}}\frac{1}{(q+\ell+k)_{+}}\bigg[\frac{1}{p_{+}}-\frac{1}{(p+q+k+\ell)_{+}}\bigg]
\nonumber\\&+&\frac{1}{(\ell
q)_{+}}\frac{1}{(k+\ell)_{+}}\frac{1}{(k-q)_{+}}\frac{1}{(\ell+k-q)_{+}}\bigg[\frac{1}{(p+k+\ell)_{+}}-\frac{1}{(p+q)_{+}}\bigg]
\nonumber\\&+&\frac{1}{(k\ell)_{+}}\frac{1}{(q+\ell)_{+}}\frac{1}{(k-q)_{+}}\frac{1}{(k-q-\ell)_{+}}\bigg[\frac{1}{(p+q+\ell)_{+}}
-\frac{1}{(p+k)_{+}}\bigg]\bigg\}.
\end{eqnarray}
By a similar argument concerning (\ref{simplified result of two
loop}), it is proved that ${\cal F}=0$. In the same way, the second
and the third terms in (\ref{finalthreeloop}) vanish. As a
consequence
\begin{eqnarray}
\Pi_{--}^{(3)}|_{nc}=0.
\end{eqnarray}
In view of the formula (\ref{spectrum}), it is deduced that
 the commutative Schwinger mass
 remains also untouched by noncommutativity at three-loop order. In
the next section, this calculation will be extended to all orders.
\section{All-loop noncommutative correction to the Schwinger mass}
\setcounter{equation}{0}
\par\noindent
In this section, we generalize three-loop computation to all orders
to obtain the exact mass spectrum. At $n$-loop level, there are
several $\theta$-dependent diagrams contributing to the vacuum
polarization tensor that one of them may be found in Fig. 4
for which $n$ is an odd number.\\
The general Feynman form of Fig. 4 related to the photon's vacuum
polarization at $n$-loop ($n\neq 1$) is written as
\begin{eqnarray}
\Pi_{--}^{(n,i)}|_{nc}&=&
(e^{2})^{n}\int\frac{dp_{+}dp_{-}}{(2\pi)^{2}}\frac{dk_{1+}dk_{1-}}{(2\pi)^{2}}\frac{dk_{2+}dk_{2-}}
{(2\pi)^{2}}\cdots\frac{dk_{(n-1)+}dk_{(n-1)-}}{(2\pi)^{2}}
~\frac{1}{k_{1-}^{2}k_{2-}^{2}\cdots
k_{(n-1)-}^{2}}\nonumber\\&\times&\exp\bigg[i\bigg(q\theta\sum\limits_{r=1}^{n-1}k_{r}
+\sum\limits_{r=1}^{\frac{n-1}{2}}k_{r}\theta\sum
\limits_{s=\frac{n+1}{2}}^{n-1}k_{s}\bigg)\bigg]\underbrace{g^{+-}\cdots g^{+-}}_{n-1}\nonumber\\
&\times &\frac{tr\bigg(\gamma_{-}
(\displaystyle{\not}q+\displaystyle{\not}p)\gamma_{-}(\displaystyle{\not}q+\displaystyle{\not}p+\displaystyle{\not}k_{1})
\cdots\gamma_{-}(\displaystyle{\not}q+\displaystyle{\not}p+\sum\limits_{i
=
1}^{n-1}\displaystyle{\not}k_{i})\gamma_{-}(\displaystyle{\not}p+\sum\limits_{i
= 1}^{n-1}\displaystyle{\not}k_{i})\cdots
\gamma_{-}\displaystyle{\not}p
\bigg)}{(q+p)^{2}(q+p+k_{1})^{2}\cdots(q+p+\sum\limits_{i=1}^{n-1}k_{i})^{2}
(p+\sum\limits_{i=1}^{n-1}k_{i})^{2}\cdots p^{2}},\nonumber\\
\end{eqnarray}
where $\Pi_{--}^{(n,i)}|_{nc}$ shows the noncommutative contribution
of the $i$th graph to the total self-energy at $n$-loop level.
Analogous to Sect. III, the numerator can be easily computed as
\begin{eqnarray}\label{nc-nloop}
\Pi_{--}^{(n,i)}|_{nc}&=&2^{1-n}
(e^{2})^{n}\int\frac{dp_{+}dp_{-}}{(2\pi)^{2}}\frac{dk_{1+}dk_{1-}}{(2\pi)^{2}}\frac{dk_{2+}dk_{2-}}
{(2\pi)^{2}}\cdots\frac{dk_{(n-1)+}dk_{(n-1)-}}{(2\pi)^{2}}
~\frac{1}{k_{1-}^{2}k_{2-}^{2}\cdots
k_{(n-1)-}^{2}}\nonumber\\&\times&\exp\bigg[i\bigg(q\theta\sum\limits_{r=1}^{n-1}k_{r}
+\sum\limits_{r=1}^{\frac{n-1}{2}}k_{r}\theta\sum
\limits_{s=\frac{n+1}{2}}^{n-1}k_{s}\bigg)\bigg]\nonumber\\
&\times
&\frac{2^{2n}~(q+p)_{-}(q+p+k_{1})_{-}\cdots(q+p+\sum\limits_{i=1}^{n-1}k_{i})_{-}
  (p+\sum\limits_{i=1}^{n-1}k_{i})_{-}\cdots p_{-}}{(q+p)^{2}(q+p+k_{1})^{2}\cdots(q+p+\sum\limits_{i=1}^{n-1}k_{i})^{2}
(p+\sum\limits_{i=1}^{n-1}k_{i})^{2}\cdots p^{2}}.
\end{eqnarray}
Due to $p$-independence of the phase factor, (\ref{nc-nloop}) can be
reduced to the following
\begin{eqnarray}
\Pi_{--}^{(n,i)}|_{nc}&=&
2^{n+1}e^{2n}\int\frac{dk_{1+}dk_{1-}}{(2\pi)^{2}}\frac{dk_{2+}dk_{2-}}
{(2\pi)^{2}}\cdots\frac{dk_{(n-1)+}dk_{(n-1)-}}{(2\pi)^{2}}
~\frac{1}{k_{1-}^{2}k_{2-}^{2}\cdots
k_{(n-1)-}^{2}}\nonumber\\&\times&
\exp\bigg[i\bigg(q\theta\sum\limits_{r=1}^{n-1}k_{r}
+\sum\limits_{r=1}^{\frac{n-1}{2}}k_{r}\theta\sum
\limits_{s=\frac{n+1}{2}}^{n-1}k_{s}\bigg)\bigg]{\cal G},
\end{eqnarray}
and $\cal{G}$ is defined as
\begin{eqnarray}\label{G-term}
{\cal G}=\int\frac{dp_{-}}{2\pi}
\frac{dp_{+}}{2\pi}\frac{1}{(q+p)_{+}(q+p+k_{1})_{+}(p+q+k_{1}+k_{2})_{+}\cdots(q+p+\sum\limits_{i=1}^{n-1}k_{i})_{+}
  (p+\sum\limits_{i=1}^{n-1}k_{i})_{+}\cdots p_{+}}.\nonumber\\
\end{eqnarray}
It is proved that for a fixed $n$, similar to the previous section,
the fraction in (\ref{G-term}) can be decomposed into partial
fractions such that
 leads to ${\cal G}=0$. Thus
\begin{eqnarray}
\Pi_{--}^{(n,i)}|_{nc}=0.
\end{eqnarray}
The obtained result is correct for any $\theta$-dependent graph.
Therefore, we conclude that
\begin{eqnarray}
\sum\limits_{i}\Pi_{--}^{(n,i)}|_{nc}=0.
\end{eqnarray}
 Accordingly, the
 noncommutativity does not affect the Schwinger mass at all orders.
\begin{figure}[h]
\vspace{-0.3cm} \centering
  \includegraphics[width=9cm,height=6cm]{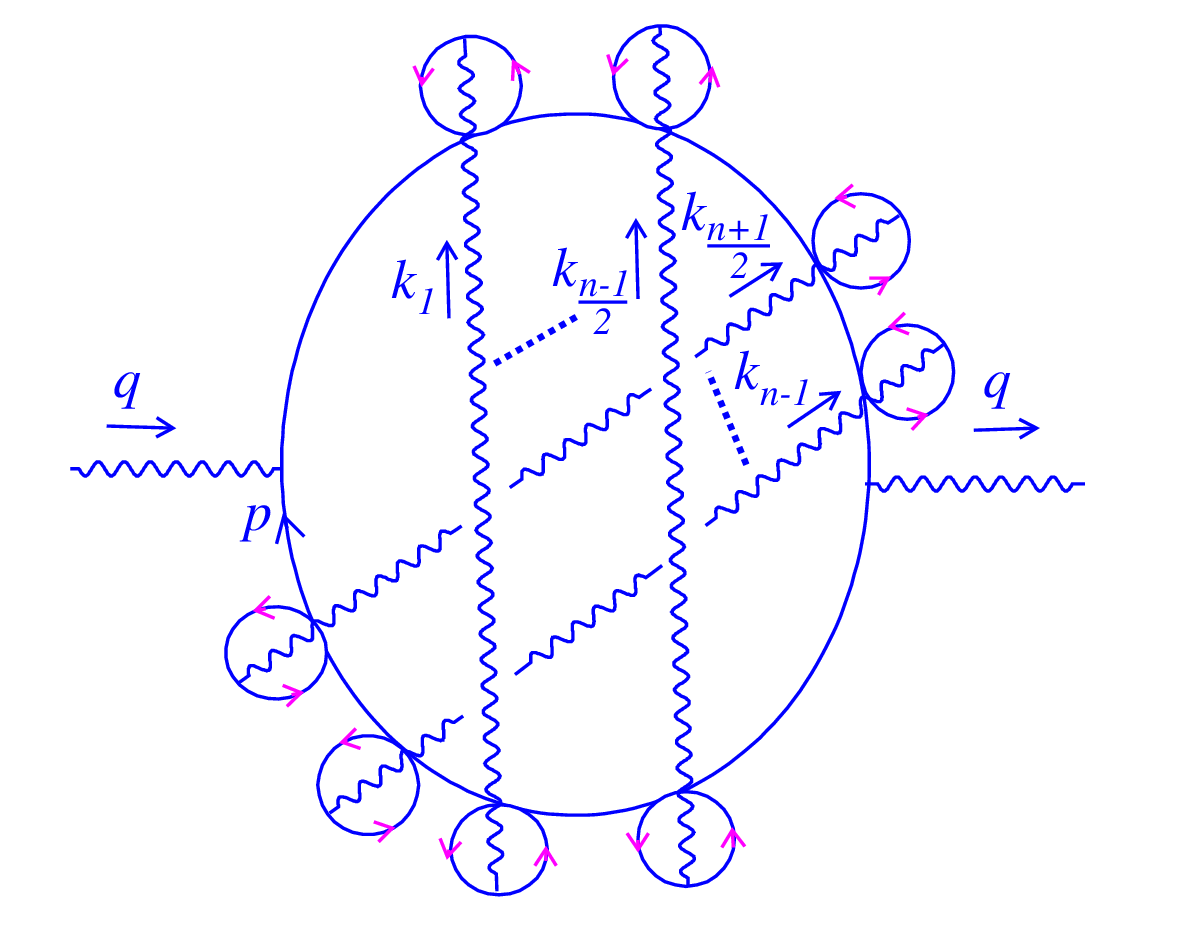}
  \caption{Relevant $n$-loop diagram
}\label{nonplanar}
\end{figure}
\\In particular, we note that diagrams like those shown in Fig. 5 with
fermionic-loop insertion produce the noncommutative phase
factors\footnote{The noncommutative phase factors related to the
graphs (a) and (b) in Fig. 5 are $e^{-i(\ell\theta q)}$ and
$e^{-i(\ell\theta q+s\theta q+s\theta\ell)}$, respectively.} which
are independent of the external fermionic-loop momentum. Hence, the
evaluation of the integral over $p$ for these graphs will be similar
to that of the graphs without the internal fermionic loops.
Consequently, it is easily shown that the contribution of these
graphs to the spectrum is also zero.
\begin{figure}[h]
 \vspace{0.03cm}
\centering
  \includegraphics[width=9cm,height=6cm]{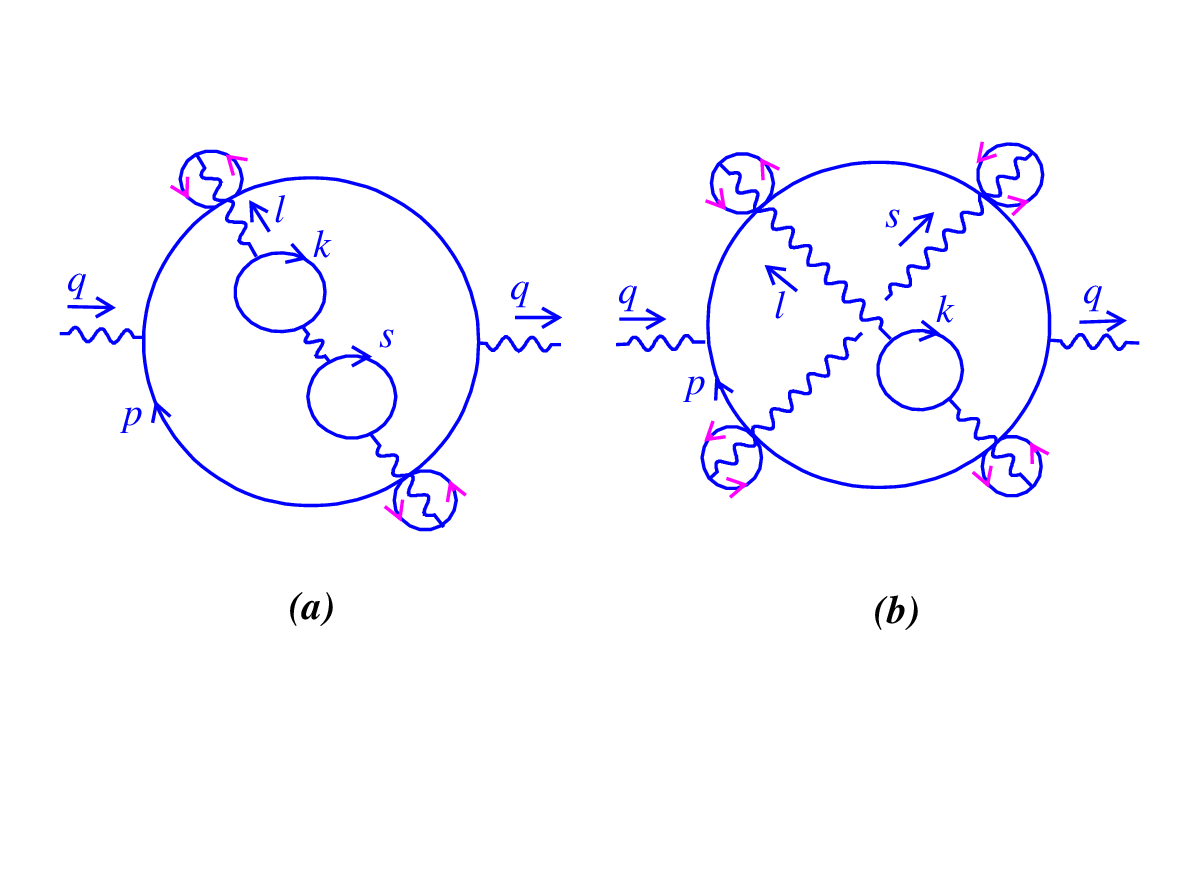}
     \vspace{-2cm} \centering
     \caption{Relevant loop diagrams with internal fermionic loop
   insertion
}\label{FeynmanRules}
\end{figure}
\section{Noncommutative one-loop effective action}
\setcounter{equation}{0}
\par\noindent
The computation method used in two previous sections will be useful
to simplify the photon's one-loop effective action in the
noncommutative space. The one-loop effective action in the
commutative space, $\Gamma^{c}[A]$, is given by integrating out the
fermionic degrees of freedom,
\begin{eqnarray}\label{gamma}
\Gamma^{c}[A]\equiv\int {\cal D}\bar\psi{\cal D}\psi\exp\bigg[i\int
d^{2}x~\bar\psi~i\displaystyle{\not}D\psi\bigg],
\end{eqnarray}
where $D_{\mu}=\partial_{\mu}-ieA_{\mu}$ and $A_{\mu}$ is an
external abelian gauge field. The quantity $\Gamma^{c}[A]$ is
equivalent to the following functional determinant from Fig. 6
\begin{eqnarray}\label{effective}
\Gamma^{c}[A]\equiv
\ln\frac{\det(\displaystyle{\not}\partial-ie\displaystyle{\not}A)}{\det(\displaystyle{\not}\partial)}=-\sum_{n=1}^{\infty}\frac{1}{n}
Tr[\frac{1}{\displaystyle{\not}\partial}(ie\displaystyle{\not}A)]^{n}.
\end{eqnarray}
\begin{figure}[h]
\vspace{-0.5cm} \centering
  \includegraphics[width=8.8cm,height=6cm]{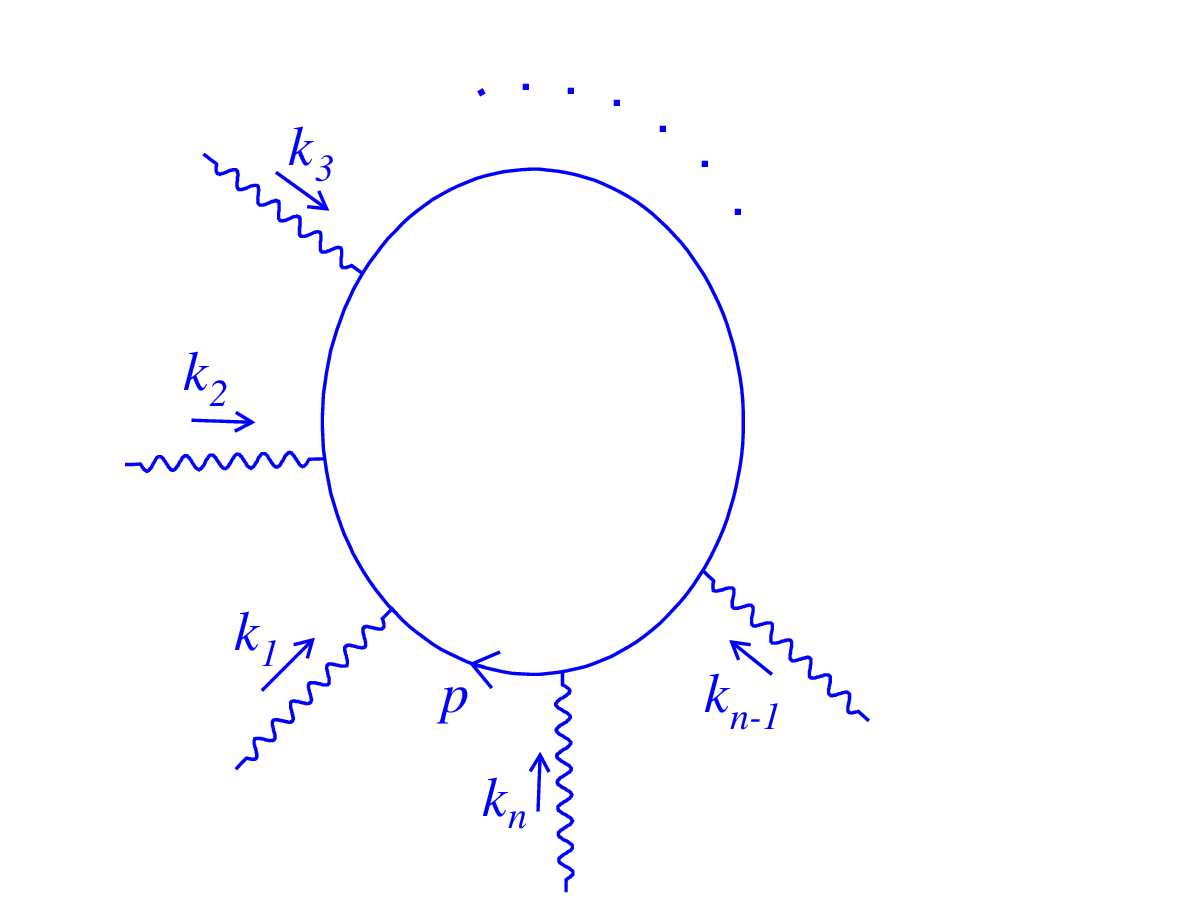}
    \caption{Relevant graph for the $n$th term of the one-loop effective
  action
}\label{nonplanar}
\end{figure}
\\
Using the non-perturbative approach in two dimensions, the
expression $\Gamma[A]$ is exactly determined. In other words,
(\ref{effective}) has non-zero value only for $n=2$ which is equal
to
\begin{eqnarray}\label{commutative value of the effective action}
\Gamma^{c}[A]=-\frac{e^{2}}{2\pi}\int
\frac{d^{2}k}{(2\pi)^{2}}A_{\mu}(k)(\delta^{\mu\nu}-\frac{k^{\mu}k^{\nu}}{k^{2}})A_{\nu}(-k).
\end{eqnarray}
Therefore, the photon has received mass from
the one-loop quantum correction. \\
The noncommutative version of $\Gamma[A]$ in three dimensions for
non-abelian gauge fields has been already discussed in
\cite{banerjee}. In what follows, we determine the one-loop
effective action for the noncommutative Schwinger model. According
to (\ref{gamma}), we can define
\begin{eqnarray}
\Gamma^{nc}[A]\equiv\int {\cal D}\bar\psi{\cal D}\psi\exp\bigg[i\int
d^{2}x~\bar\psi\star~i\displaystyle{\not}D\psi\bigg],
\end{eqnarray}
where $D_{\mu}=\partial_{\mu}-ieA_{\mu}\star$. Similar to the
commutative part, $\Gamma^{nc}[A]$ can be represented as
\begin{eqnarray}
\Gamma^{nc}[A]=\ln\frac{\det(\displaystyle{\not}\partial-ie\displaystyle{\not}A\star)}{\det(\displaystyle{\not}\partial)}=-\sum_{n=1}^{\infty}\frac{1}{n}
Tr[\frac{1}{\displaystyle{\not}\partial}(ie\displaystyle{\not}A\star)]^{n},
\end{eqnarray}
which is equivalent to the following expression:
\begin{eqnarray}
\Gamma^{nc}[A]=\sum_{n=1}^{\infty}\int d^{2}z_{1}\cdots
d^{2}z_{n}~A^{\mu_{1}}(z_{1})\cdots
A^{\mu_{n}}(z_{n})~\Gamma_{\mu_{1}\cdots\mu_{n}}^{nc}(z_{1},\cdots,z_{n}).
\end{eqnarray}
The quantity
$\Gamma_{\mu_{1}\cdots\mu_{n}}^{nc}(z_{1},\cdots,z_{n})$ is given
by\footnote{In the commutative case, $\theta=0$, only
$\Gamma_{\mu_{1}\mu_{2}}^{c}$ has non-zero value and
$\Gamma_{\mu_{1}\cdots\mu_{n}}^{c}$ vanishes for $n>2$.}
\begin{eqnarray}\label{nc-effective action at 1-loop}
\Gamma_{\mu_{1}\cdots\mu_{n}}^{nc}(z_{1},\cdots,z_{n})&=&\frac{(-e)^{n}}{n}\int\prod_{j=1}^{n}\frac{d^{2}k_{j}}{(2\pi)^{2}}~(2\pi)^{2}
\delta(\sum_{j=1}^{n}k_{j})~
e^{i\sum\limits_{j=1}^{n}k_{j}z_{j}}e^{\frac{i}{2}\sum\limits_{j<\ell}^{n}k_{j}\theta
k_{\ell}}\widetilde{\Gamma}_{\mu_{1}\cdots\mu_{n}}(k_{1},\cdots,k_{n}),\nonumber\\
\end{eqnarray}
with
\begin{eqnarray}\label{integral over fermion loop}
\widetilde{\Gamma}_{\mu_{1}\cdots\mu_{n}}(k_{1},\cdots,k_{n})=
\int\frac{d^{2}p}{(2\pi)^{2}}~\frac{tr\bigg(\gamma_{\mu_{1}}
(\displaystyle{\not}p+\displaystyle{\not}k_{1})
\gamma_{\mu_{2}}(\displaystyle{\not}p+\displaystyle{\not}k_{1}+\displaystyle{\not}k_{2})
\gamma_{\mu_{3}}(\displaystyle{\not}p+\displaystyle{\not}k_{1}+\displaystyle{\not}k_{2}+\displaystyle{\not}k_{3})\cdots
\gamma_{\mu_{n}}\displaystyle{\not}p\bigg)
}{(p+k_{1})^{2}(p+k_{1}+k_{2})^{2}(p+k_{1}+k_{2}+k_{3})^{2}\cdots p^{2}}.\nonumber\\
\end{eqnarray}
Since the noncommutative phase factor produced in (\ref{nc-effective
action at 1-loop}), similar to Sect. III and IV, is also
$p$-independent, the integral over $p$ can be separated from the
rest, i.e. (\ref{integral over fermion loop}). \\The non-zero
leading term in (\ref{nc-effective action at 1-loop}) arises from
$n=2$ which leads to its commutative value\footnote{Since for $n=2$
the noncommutative phase factors arising from two vertices cancel
each other, we obtain the commutative result (\ref{commutative value
of the effective action}).}, namely
$\Gamma_{\mu_{1}\mu_{2}}^{nc}=\Gamma_{\mu_{1}\mu_{2}}^{c}$. For
$n>2$, we just follow the technique applied for two- and three-loop
calculations. Writing (\ref{integral over fermion loop}) in the
light-cone coordinates, we arrive at
\begin{eqnarray}\label{integral over fermion loop in light-cone}
\widetilde{\Gamma}_{-\cdots-}(k_{1},\cdots,k_{n})=
\int\frac{dp_{-}}{(2\pi)}\frac{dp_{+}}{(2\pi)}~\frac{tr\bigg(\gamma_{-}
(\displaystyle{\not}p+\displaystyle{\not}k_{1})
\gamma_{-}(\displaystyle{\not}p+\displaystyle{\not}k_{1}+\displaystyle{\not}k_{2})
\gamma_{-}(\displaystyle{\not}p+\displaystyle{\not}k_{1}+\displaystyle{\not}k_{2}+\displaystyle{\not}k_{3})\cdots
\gamma_{-}\displaystyle{\not}p\bigg)
}{(p+k_{1})^{2}(p+k_{1}+k_{2})^{2}(p+k_{1}+k_{2}+k_{3})^{2}\cdots p^{2}}.\nonumber\\
\end{eqnarray}
Using the detailed computations of Appendix A, (\ref{integral over
fermion loop in light-cone}) can be simplified as
\begin{eqnarray}\label{simplisied gamma}
\widetilde{\Gamma}_{-\cdots-}(k_{1},\cdots,k_{n})=
\int\frac{dp_{-}}{(2\pi)}\frac{dp_{+}}{(2\pi)}~\frac{1
}{(p+k_{1})_{+}(p+k_{1}+k_{2})_{+}(p+k_{1}+k_{2}+k_{3})_{+}...
p_{+}}.\nonumber\\
\end{eqnarray}
Analogous to (\ref{G-term}), the relation (\ref{simplisied gamma})
can be decomposed into partial fractions for a fixed $n$. After
doing complete decomposition and using the complex form of Green's
theorem, we obtain $\Gamma_{\mu_{1}\cdots\mu_{n}}^{nc}=0$ for $n>2$.
Thus, the noncommutativity has no effect on one-loop effective
action and its exact commutative form is preserved.
\begin{eqnarray}
\Gamma^{nc}[A]=\Gamma^{c}[A]=-\frac{e^{2}}{2\pi}\int
\frac{d^{2}k}{(2\pi)^{2}}A_{\mu}(k)(\delta^{\mu\nu}-\frac{k^{\mu}k^{\nu}}{k^{2}})A_{\nu}(-k).
\end{eqnarray}
\section{Conclusion}
\noindent
In this paper, we have concentrated on the mass spectrum of the
noncommutative Schwinger model with Euclidean signature at higher
loops. It is demonstrated that the Schwinger mass receives no
noncommutative corrections at all orders.
\par
To prove this in a
perturbative method, we have used the light-cone gauge to simplify
the Lagrangian form. In this gauge, only the fermion-photon vertex
remains and consequently the fermionic loops contribute to our
calculations. Having fixed the gauge, the study of the
noncommutative sector of the photon self-energy at two- , three-,
and all-loop order has been performed. \par
 At two- and
three-loop level, the noncommutative parts of the photon self-energy
were analyzed. Since the noncommutative phase factor appearing in
Feynman integrals is independent of the fermionic-loop momentum, the
corresponding loop integral is easily evaluated. This analysis
showed that the contributions from the $\theta$-dependent graphs are
zero. Hence, the commutative mass spectrum does not change at these
orders. Then, the calculation of Sect. III was extended to all
orders. Similar to two- and three-loop level, the noncommutative
phase factor is independent of the fermionic-loop momentum and the
resulting integral vanishes. This proves that the Schwinger mass
remains intact at all orders in the noncommutative space.
\par
 The
technique applied for computing the trace of the fermionic loops
inspired us to study the relevant one-loop effective action. As a
consequence, the exact one-loop effective action in the light-cone
gauge with no noncommutative corrections was obtained. \\
Using the arguments of Sects. III and IV, it is possible to
extend the analysis of the one-loop effective action to all loops.
It is easily shown that the all-loop photon's effective action,
similar to one-loop effective action, does not also receive
noncommutative corrections. Although we have investigated
in this paper only the
photon sector, it would be interesting to do a similar analysis for
the fermion self-energy and running of the coupling constant, in which case noncommutativity corrections are expected to appear.
\section{Acknowledgments}
\par\noindent
I would like to express my special gratitude to F. Ardalan for his inspiration and
illuminating discussions. Special thanks go to M.M. Sheikh-Jabbari, M. Khorrami and A.A. Varshovi for
critical remarks and fruitful conversation; and also I appreciate the useful comments of A. Armoni.
\appendix
\section{Two-loop fermionic trace in the light-cone coordinates}
\noindent
In this appendix, we present more details of the computation of the
trace expression appearing in the relation (\ref{2loopnum}).
\begin{eqnarray}\label{numerator-1}
&&{\cal N}=tr\bigg(\gamma_{-}
(\displaystyle{\not}q+\displaystyle{\not}p)\gamma_{-}(\displaystyle{\not}q+
\displaystyle{\not}p+\displaystyle{\not}k)\gamma_{-}(\displaystyle{\not}p+\displaystyle{\not}k)
\gamma_{-} \displaystyle{\not}p \bigg).
\end{eqnarray}
To calculate this, we start from the representation of the gamma
matrices in Euclidean space
  \begin{eqnarray}
\gamma_{1}=\left(
  \begin{array}{cc}
    0 & -i \\
    i & 0 \\
  \end{array}
\right),~~~~\gamma_{2}=\left(
  \begin{array}{cc}
    0 & -1 \\
    -1 & 0 \\
  \end{array}\right),
  \end{eqnarray}
 which in the light-cone coordinates are defined as
  \begin{eqnarray}
\gamma_{+}=\gamma_{1}+i\gamma_{2}= \left(
  \begin{array}{cc}
    0 & -2i\\
   0 & 0 \\
  \end{array}\right)
,~~~~ \gamma_{-}=\gamma_{1}-i\gamma_{2}= \left(
  \begin{array}{cc}
    0 & 0\\
   2i & 0 \\
  \end{array}\right),
  \end{eqnarray}
and the light-cone metric by using $g^{\mu\nu}=\frac{\partial
x^{\mu}}{\partial x^{\rho}}\frac{\partial x^{\nu}}{\partial
x^{\sigma}}\delta^{\rho\sigma}$ is obtained
\begin{eqnarray}
g^{\mu\nu}=\left(
  \begin{array}{cc}
    g^{++} & g^{+-}\\
   g^{-+} & g^{--} \\
  \end{array}\right)=
  \left(
  \begin{array}{cc}
    0 & \frac{1}{2}\\
 \frac{1}{2} & 0 \\
  \end{array}\right).
\end{eqnarray}
The terms such as $\displaystyle{\not}p$ appeared in
(\ref{numerator-1}) can be revised as the following
\begin{eqnarray}\label{p in lightcone}
\displaystyle{\not}p=\frac{1}{2}(p_{+}\gamma_{-}+p_{-}\gamma_{+})=
  \left(
  \begin{array}{cc}
    0 & -ip_{-}\\
  ip_{+} & 0 \\
  \end{array}\right),
\end{eqnarray}
consequently
\begin{eqnarray}\label{gamma p}
\gamma_{-}\displaystyle{\not}p=\left(
  \begin{array}{cc}
    0 & 0\\
   2i & 0 \\
  \end{array}\right)\left(
  \begin{array}{cc}
    0 & -ip_{-}\\
  ip_{+} & 0 \\
  \end{array}\right)=\left(
  \begin{array}{cc}
    0 & 0\\
  0& 2p_{-} \\
  \end{array}\right).
\end{eqnarray}
Plugging (\ref{gamma p}) in (\ref{numerator-1})
\begin{eqnarray}\label{calc trace}
{\cal N}&=&tr\bigg[ \left(
  \begin{array}{cc}
    0 & 0\\
  0& 2(p_{-}+q_{-}) \\
  \end{array}\right)\left(
  \begin{array}{cc}
    0 & 0\\
  0& 2(p+q+k)_{-} \\
  \end{array}\right)\left(
  \begin{array}{cc}
    0 & 0\\
  0& 2(p+k)_{-} \\
  \end{array}\right)
  \left(
  \begin{array}{cc}
    0 & 0\\
  0& 2p_{-} \\
  \end{array}\right)
\bigg] \nonumber\\&=&tr\left(
  \begin{array}{cc}
    0 & 0\\
  0& 2^{4}(p+q)_{-}(p+q+k)_{-}(p+k)_{-}p_{-} \\
  \end{array}\right)
\nonumber\\&=&2^{4}~(p+q)_{-}(p+q+k)_{-}(p+k)_{-}p_{-}.
\end{eqnarray}
\section{Two-loop photon self-energy with mass insertion}
\noindent
Our purpose of the present appendix is to illustrate that the
electron mass insertion as an infrared regulator does not change the
result (\ref{2loopresult}). To prove this, let us start from the
relation (\ref{2loop self}) by rewriting it with the mass term
\begin{eqnarray}\label{self with mass}
\Pi_{--}^{(2)}|_{nc}&=&
e^{4}\int\frac{dp_{-}dp_{+}}{(2\pi)^{2}}\frac{dk_{+}dk_{-}}{(2\pi)^{2}}
~g^{+-}e^{-ik\theta q}{\cal N}_{_{m}} \nonumber\\&\times&\frac{1}
{k^{2}_{-}[(q+p)^{2}-m^{2}][(q+p+k)^{2}-m^{2}][(p+k)^{2}-m^{2}][p^{2}-m^{2}]},
\end{eqnarray}
where
\begin{eqnarray}\label{numerator with m}
{\cal N}_{_{m}}=tr\bigg(\gamma_{-}
(\displaystyle{\not}q+\displaystyle{\not}p+m)\gamma_{-}(\displaystyle{\not}q+
\displaystyle{\not}p+\displaystyle{\not}k+m)\gamma_{-}(\displaystyle{\not}p+\displaystyle{\not}k+m)
\gamma_{-} (\displaystyle{\not}p+m) \bigg).
\end{eqnarray}
Similar to (\ref{p in lightcone}), the matrix form of
$\displaystyle{\not}p+m$ in light-cone coordinates is given by
\begin{eqnarray}
\displaystyle{\not}p+m=
  \left(
  \begin{array}{cc}
    m & -ip_{-}\\
  ip_{+} & m \\
  \end{array}\right),
\end{eqnarray}
multiplying by $\gamma_{-}$, we have
\begin{eqnarray}\label{matrix form of p+m}
\gamma_{-}(\displaystyle{\not}p+m)=\left(
  \begin{array}{cc}
    0 & 0\\
   2i & 0 \\
  \end{array}\right)\left(
  \begin{array}{cc}
    m & -ip_{-}\\
  ip_{+} & m \\
  \end{array}\right)=\left(
  \begin{array}{cc}
    0 & 0\\
  2im& 2p_{-} \\
  \end{array}\right).
\end{eqnarray}
Substituting (\ref{matrix form of p+m}) in (\ref{numerator with m})
 yields
\begin{eqnarray}\label{calc trace with m}
{\cal N}_{_{m}}&=&tr\bigg[ \left(
  \begin{array}{cc}
    0 & 0\\
  2im& 2(p_{-}+q_{-}) \\
  \end{array}\right)\left(
  \begin{array}{cc}
    0 & 0\\
  2im& 2(p+q+k)_{-} \\
  \end{array}\right)\left(
  \begin{array}{cc}
    0 & 0\\
  2im& 2(p+k)_{-} \\
  \end{array}\right)
     \left(
    \begin{array}{cc}
    0 & 0\\
  2im& 2p_{-} \\
  \end{array}\right)
\bigg]\nonumber\\& =&16~tr\left(
  \begin{array}{cc}
    0 & 0\\
  im {\cal J}& {\cal J}\\
  \end{array}\right)
=16{\cal J},
\end{eqnarray}
with
\begin{equation}
{\cal J}=(p+q)_{-}(p+q+k)_{-}(p+k)_{-}p_{-}.
\end{equation}
As we see, the mass of the electron does not appear in the final
result of the trace expression. Hence, the relations (\ref{calc
trace with m}) and (\ref{calc trace}), corresponding to ${\cal
N}_{_{m}}$ and ${\cal N}$, respectively, are exactly the same, apart
from a numerical factor.\\ Inserting (\ref{calc trace with m}) in
(\ref{self with mass}) and simplifying, we arrive at
\begin{eqnarray}
\Pi_{--}^{(2)}|_{nc}&=&
8e^{4}\int\frac{dk_{+}}{2\pi}\frac{dk_{-}}{2\pi} ~\frac{e^{-ik\theta
q}}{k^{2}_{-}}~{\cal E}_{_{m}},
\end{eqnarray}
where
\begin{eqnarray}\label{E with m}
{\cal E}_{_{m}}=\int\frac{dp_{-}}{2\pi}\frac{dp_{+}}{2\pi}
~\frac{1}{[(p+q)_{+}-\frac{m^{2}}{(p+q)_{-}}]~
[(p+q+k)_{+}-\frac{m^{2}}{(p+q+k)_{-}}]~[(p+k)_{+}-\frac{m^{2}}{(p+k)_{-}}]~[p_{+}-\frac{m^{2}}{p_{-}}]}.\nonumber\\
\end{eqnarray}
Decomposing the integrand of (\ref{E with m}) into partial fractions, we have
\begin{eqnarray}
{\cal E}_{_{m}}&=&\int\frac{dp_{-}}{2\pi}\frac{dp_{+}}{2\pi}
~\bigg(\frac{1}{k_{+}+\frac{m^{2}k_{-}}{(p+q)_{-}(p+q+k)_{-}}}\bigg)
\bigg(\frac{1}{(p+q)_{+}-\frac{m^{2}}{(p+q)_{-}}}
-\frac{1}{(p+q+k)_{+}-\frac{m^{2}}{(p+q+k)_{-}}}\bigg)
\nonumber\\&\times&\bigg(\frac{1}{k_{+}+\frac{m^{2}k_{-}}{p_{-}(p+k)_{-}}}\bigg)
\bigg(\frac{1}{p_{+}-\frac{m^{2}}{p_{-}}}-\frac{1}{(p+k)_{+}-\frac{m^{2}}{(p+k)_{-}}}\bigg),
\end{eqnarray}
which leads to the final result as
\begin{eqnarray}
{\cal E}_{_{m}}&=&\int\frac{dp_{-}}{2\pi}\frac{dp_{+}}{2\pi}
~\bigg(\frac{1}{k_{+}+\frac{m^{2}k_{-}}{(p+q)_{-}(p+q+k)_{-}}}\bigg)\bigg(\frac{1}{k_{+}+\frac{m^{2}k_{-}}{p_{-}(p+k)_{-}}}\bigg)
\nonumber\\&\times&\bigg\{\bigg(\frac{1}{q_{+}+\frac{m^{2}q_{-}}{q_{-}(p+q)_{-}}}\bigg)\bigg[\frac{1}{p_{+}-\frac{m^{2}}{p_{-}}}
-\frac{1}{(p+q)_{+}-\frac{m^{2}}{(p+q)_{-}}}\bigg]\nonumber\\&-&
\bigg(\frac{1}{(k-q)_{+}+\frac{m^{2}(k-q)_{-}}{(p+q)_{-}(p+k)_{-}}}\bigg)\bigg[\frac{1}{(p+q)_{+}-\frac{m^{2}}{(p+q)_{-}}}
-\frac{1}{(p+k)_{+}-\frac{m^{2}}{(p+k)_{-}}}\bigg] \nonumber\\&-&
\bigg(\frac{1}{(k+q)_{+}+\frac{m^{2}(k+q)_{-}}{p_{-}(p+q+k)_{-}}}\bigg)\bigg[\frac{1}{p_{+}-\frac{m^{2}}{p_{-}}}
-\frac{1}{(p+q+k)_{+}-\frac{m^{2}}{(p+q+k)_{-}}}\bigg]
\nonumber\\&+&
\bigg(\frac{1}{q_{+}+\frac{m^{2}q_{-}}{(p+k)_{-}(p+q+k)_{-}}}\bigg)
\bigg[\frac{1}{(p+k)_{+}-\frac{m^{2}}{(p+k)_{-}}}-\frac{1}{(p+q+k)_{+}-\frac{m^{2}}{(p+q+k)_{-}}}\bigg]\bigg\}.\nonumber\\
\end{eqnarray}
Using the complex version of Green's theorem yields ${\cal
E}_{_{m}}=0$ and consequently $\Pi_{--}^{(2)}|_{nc}=0$.

\end{document}